\documentclass{iaus}
\usepackage{graphicx}

\title[Separating Physical Components from Spectra] 
{Separating Physical Components from Galaxy Spectra by Subspace Methods}

\author[C.~W.~Yip et al.]   
{Ching-Wa~Yip$^{1}$\thanks{Keck Fellow. Email to authors:
  cwyip@pha.jhu.edu; szalay@pha.jhu.edu}, Alex~S.~Szalay$^1$, 
  Andrew~J.~Connolly$^2$ and Tamas~Budav\'ari$^1$}

\affiliation{$^1$Department of Physics and Astronomy, The Johns
  Hopkins University, Baltimore, MD 21218, USA  \\[\affilskip]
$^2$Department of Physics and Astronomy, University of Pittsburgh,
  Pittsburgh, PA 15260, USA \\[\affilskip]}
\pubyear{2006}
\volume{241} 
\pagerange{xxx}
\date{?? and in revised form ??}
\jname{Stellar Populations as Building Blocks of Galaxies}
\editors{R.~F.~Peletier, A.~Vazdekis, eds.}
\begin{document}

\maketitle

\begin{abstract}
Using  subspace  methods,  we   study  the  distribution  of  physical
components of galaxies  in wavelength space. We find  that it is valid
to assume that the stellar and the gaseous components of galaxies span
complementary subspaces. To first  order, stellar and gaseous spectral
features can be extracted from  galaxy spectra through a simple matrix
multiplication.    By   comparing   the  stellar   continua   obtained
respectively using the model-based  and the empirical approach through
a  commonality   measure,  we  conclude  that  the   latter  may  lose
higher-order   spectral   features.  
\end{abstract}

\firstsection 
\section{Introduction}

A common  method of estimating  physical parameters of galaxies  is by
least-square  fitting  stellar population  synthesis  models to  their
spectra.  The approach is time-consuming when the model becomes large.
This hinders  the examination of all possible  parameters defining the
models (for example,  P\'EGASE \cite{pegase} is defined by  as much as
$\sim$10  parameters), especially for  large sky  surveys such  as the
SDSS which will produce 1  million galaxy spectra.  On the other hand,
principal  components (PCs) are  known to  be powerful  in classifying
galaxy spectra \cite{Connolly95}.  This suggests that, as PCs form one
kind of basis vectors, a subspace  is a good description to the higher
dimensional  spectral space.  In  this work,  we exploit  the subspace
methods in extracting physical information from galaxy spectra.

\section{Stellar and Gaseous Subspaces}

\begin{figure}[ht]
\begin{minipage}[t]{6.5cm}
\begin{center}
\includegraphics[width=5.cm,clip]{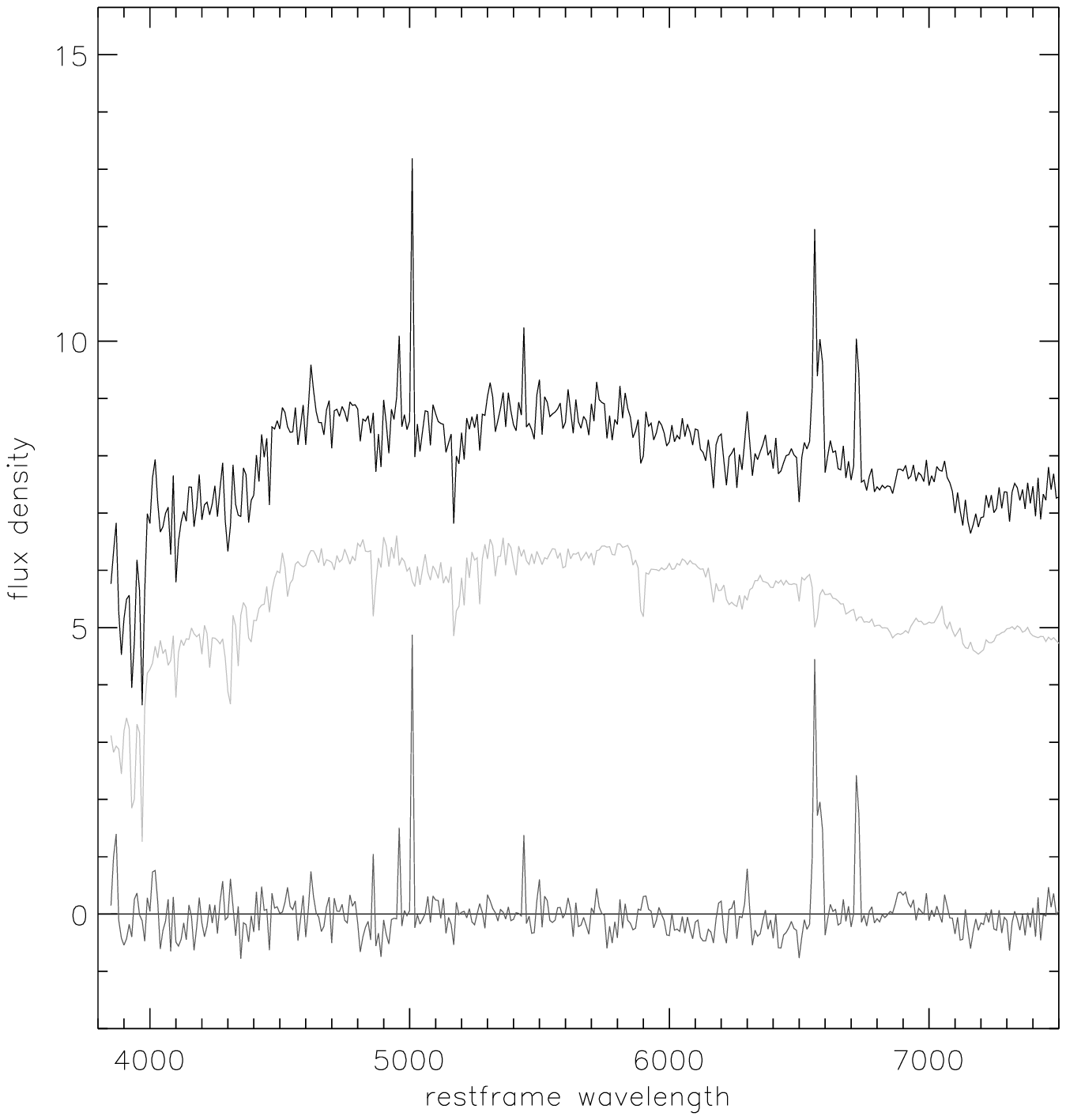}
\caption{The stellar (light gray) and gaseous (dark gray) components
of the observed galaxy spectrum (black) obtained by the subspace method. 
The flux  density of the  stellar component is
arbitrarily offset for clarity.}
\label{fig:yip_1_fig1}
\end{center}
\end{minipage}
\hfill
\begin{minipage}[t]{6.5cm}
\begin{center}
\includegraphics[width=5.cm,clip]{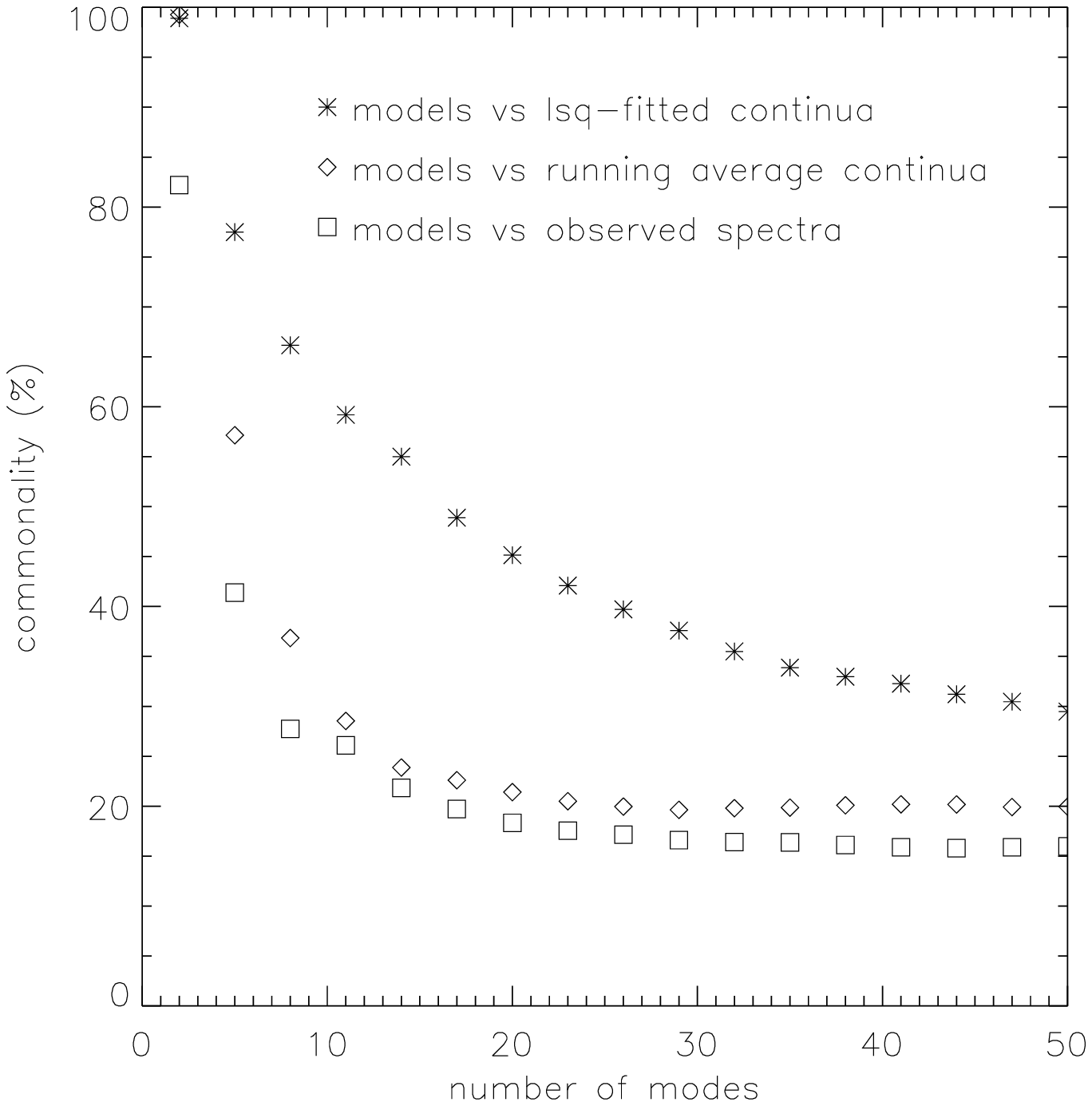}
\caption[Short caption for figure 2]{The commonality between the
    stellar subspaces obtained by different methods, as a function of the 
    number of eigenspectra in each and all of the sets 
    of spectra. }
\label{fig:yip_1_fig2} 
\end{center}
\end{minipage}
\end{figure}

The projection  matrix for a set of  astronomical spectra \cite{Yip04}
is defined to  be $P \equiv \sum_{e} |e\rangle  \langle e|$, where $e$
are  the eigenspectra \cite{Connolly95}  of that  set of  spectra.  We
construct  a projection  matrix using  the \cite{BC03}.   The subspace
defined by  this projection  matrix, $P(\mbox{star})$, is  referred as
the stellar subspace.   Each galaxy spectrum $f$ in  our sample (taken
to be  various types  of SDSS galaxy  spectra) is projected  onto this
stellar subspace, so that the resultant vector, $f(\mbox{star})$, is
\begin{equation}
f(\mbox{star}) \equiv P(\mbox{star}) \times f \ ,
\end{equation}

\noindent where $\times$ means a matrix multiplication.
If  the  gaseous and stellar subspaces are complementary, we have
\begin{equation}
P(\mbox{gas}) \equiv I - P(\mbox{star})  \ ,
\end{equation}

\noindent where $I$ is the  identity matrix. The gaseous component can
be obtained directly by $f(\mbox{gas}) \equiv P(\mbox{gas}) \times f$.

Fig.~\ref{fig:yip_1_fig1}   shows  the  resultant   projected  spectra
$f(\mbox{star})$ and  $f(\mbox{gas})$ for an  example galaxy spectrum,
which resemble the expected  stellar and gaseous contributions.  Since
for  a  given subspace,  its  projection  matrix  is uniquely  defined
\cite{Oja83}, only a  single projection matrix is needed  to define an
arbitrarily complex model.

\section{Definition of  Stellar Subspace: 
Model-Based vs. Empirical}

As  we  construct  the   stellar  subspace  by  adopting  the  stellar
population synthesis  models, it is  important to understand  how well
this model-based approach is when compared with other approach such as
the  empirical  approach.  Our  chosen  empirical  approach  is for  a
particular  pixel by  averaging the  flux  between the  40th and  60th
quantiles of  the flux distribution  taken from a wavelength  range of
$\pm 200$~km/s.   The emission  lines are masked  around $\lambda_{0}$
$\pm 2000$~km/s,  with $\lambda_{0}$ being the wavelength  of the line
center.

To  measure  the overlap  between  the  stellar  subspaces defined  by
different approaches, we use the commonality measure (see \cite[Yip et
al.   2004]{Yip04}  for  its application).   Fig.~\ref{fig:yip_1_fig2}
shows that  the least-square fit stellar  continua (i.e.  model-based)
do  not overlap 100\%  with the  model spectra  because of  the sample
variation arise  from the selection of  the galaxy.  This  will be our
null  measure, as such  no other  stellar continuum  estimation method
would  exceed  this  line.   The  commonality  between  the  empirical
running-average approach for the  stellar continua and the model drops
below that  of the null measure  with the number  of modes, indicating
the loss of higher-order features in the former.

We acknowledge support from the W.~M.~Keck Foundation, through a grant
given to establish a program of data intensive science at JHU.

\end{document}